\documentclass{article}
\usepackage{graphicx} 
\usepackage{amsfonts}
\usepackage{amssymb}
\usepackage{amsmath}
\usepackage{xcolor}
\usepackage{xspace}
\usepackage{pdfpages}
\usepackage{url}
\usepackage{hyperref}

\title{0xPass: A Secure Protocol for Universal Cross-Chain Accounts}
\author{Keon Kim \and Krish Chelikavada}
\date{19/02/2024}

\begin{document}

\maketitle

\section{Architecture}

The 0xPass platforms allows clients to seamlessly operate across multiple cryptocurrency ecosystems using a single account. In order to do so, it maintains user accounts mapping a user's assets in these different ecosystems and executes complex operations by moving such assets across ecosystems. In order to issue a request for performing a given operation, the client must first authenticate their identity to the 0xPass platform, which then determines the individual transactions that must be executed. Finally, 0xPass automatically signs all transactions by means of threshold signing protocols without ever storing the signing keys in any single node, in order to protect the user's assets. The 0xPass platform is composed of the following 3 layers, which interoperate in a modular fashion:

\begin{itemize}
    \item \textbf{Orchestrator Layer.} This layer maintains account metadata and handles client requests, including authentication and authorization. Upon validating a request from a user, it invokes the solver layer to determine the sequence of transactions needed to perform the request. Once this sequence of transactions is determined by the Solver layer, the Orchestration layer authorizes each transaction so that it can be executed by the Transaction layer. Alternatively, if the user chooses to delegate the power to authorize the sequence of transactions needed to fulfill a request, the Orchestration layer further delegates this power to the Solver layer\footnote{The Orchestration layer can temporarily authorize the Solver layer to order the Transaction layer to sign a single sequence of transactions. This authorization only allows the Solver layer to order the signing of one sequence of transactions during a short period of time, and is automatically revoked after this sequence of transactions is signed.}.
    
    \item \textbf{Solver Layer.} This layer determines how to perform complex requests that may involve multiple transactions across multiple ecosystems, including L-1 and L-2 blockchains and bridges. It uses account metadata about a client's different assets to determine how to better perform the requested operation and creates a sequence of transactions that must be executed on different platforms and ecosystems in order to compelte the client's request. This sequence of transactions is sent for authorization by the Orchestration layer, potentially involving consent from the user for the particular sequence of transactions. Alternatively, if the user has delegated the power for authorizing the sequence of transactions for a given request, the Solver layer directly orders the Transaction layer to sign the sequence of transactions.

    \item \textbf{Transaction Layer.} This layer uses threshold signature protocols to sign transactions moving a user's assets. In order to sign a transaction, the Transaction layer requires the transaction to be authorized by both the user and the Orchestration layer, which does so after authenticating the user and checking that the transaction obeys rules stated as part of metadata in the user's account. This layer also initially maintains signature key shares needed to execute threshold signing protocols. In later decentralization phases, this role may be delegated to a Key Management sub-layer.
\end{itemize}

\paragraph{Universal Accounts and Complex Requests.} User create universal accounts managed by the Orchestrator layer, which is responsible for storing account metadata as well as handling user authentication and transaction authorization. Once a user has created an account, they can authenticate themselves to the Orchestrator layer and issue complex requests for operations among multiple cryptocurrency platforms supported by 0xPass. Such operations are performed by a sequence of transactions determined by the Solver layer and executed by the Transaction layer. This process is invisible to the final user and developers integrating their solutions into the 0xPass platform. 

\paragraph{Modular Design with Layer Modules.} Each layer is designed in a modular fashion, allowing for new features  to be added by deploying new modules as the platform evolves. The different modules executing tasks in each layer communicate via an API that will be ultimately available to third party developers. For example, this modular architecture allows for adding new threshold signing protocols or support to new cryptocurrencies as the platform evolves.

\paragraph{Decentralized Execution with Sub-Networks.} In the first phase, the platform is deployed in a centralized manner and controlled by 0xPass. Later decentralization phases will allow third party organizations not only to deploy new modules but also to host their own sub-networks of nodes executing these modules. Different modules on each of the layers can be executed by different sub-networks while communicating and issuing requests to each other.

\paragraph{Backup Across Sub-Networks.} Besides executing modules offering new features to each layer, multiple sub-networks can also act as backup services to each other. Since modules require both public metadata (\textit{e.g.} addresses associated to an account) and sensitive private data (\textit{e.g.} key shares) to function, it is natural that this data is primarily maintained by the sub-network natively executing the module. However, in order to provide fault tolerance beyond a single sub-network, other sub-networks can provide backup storage services as decentralization progresses. 
In the context of public data, this feature will allow for easily integrating existing Data Availability (DA) services into the 0xPass platform. In the context of sensitive private data, new Key Management modules specifically designed for securely holding private data in a distributed manner can be deployed.

\paragraph{Fast Execution of Complex Requests.} Fulfilling a complex request issued by a client may require executing a long sequence of transactions across many different cryptocurrency ecosystems, which has an intrinsic delay depending on how fast each ecosystem processes transactions. While the Solver layer of the 0xPass network can take care of successfully executing such long transaction sequences without user interaction after the request is issued, waiting for all transactions to be executed will delay updating the client's balance in the universal account. In order to speed up this process, the Solver Layer can optimistically update the client's balance while executing the sequence of transactions on the background. This feature can be implemented in different ways, including explicitly locking client resources to ensure successful execution of all transactions and using the request authorization delegation feature described in Section~\ref{sec:auth} to delegate to the Solver Layer the power to execute all necessary transactions. As the decentralization plan advances, different approaches for such speed ups can be added to the 0xPass network by adding new modules potentially executed by third party sub-networks.

\section{Authentication and Authorization Across Layers}\label{sec:auth}

One of 0xPass' biggest strengths is allowing for seamless interoperability between different cryptocurrency platforms, which are controlled by a single user account. While ease of use for both users and developers is the main goal, it is also paramount to provide strong user authentication methods, in order to avoid account misuse by attackers. For a user to safely issue requests to the 0xPass platform, it must not only prove their identity to the Orchestrator Layer but also authorize requests that are passed down to further layers.

Each Layer is composed by several Modules implementing different features, including interoperability with different external platforms. In this context, it is essential to provide integration between modules that need to issue requests to each other. When authorizing a request, it is necessary to ensure that not only it is issued by a module with the necessary privilege but also that this privilege has been granted by a user.

As the decentralization plan progresses and third party organizations are allowed to deploy and execute their own modules and sub-networks, we will deploy a comprehensive user authentication and transaction authorization solution. This solution will handle both user authentication towards the Orchestrator layer and the authorization of individual transaction requests issued by the Solver layer to the Transaction layer. At the core of this solution are Single Sign-on services with support for multi-factor authentication to establish users' identities combined with cryptographic protocols for authorizing each transaction issued by the 0xPass platform.

\subsection{User Registration}

User registration and authentication are handled by the Orchestrator layer, which maintains user account metadata and communicates with third party single Sign-on providers. At registration time, a user's identity is verified and their account information is created and stored by the Orchestrator layer, so that it can be used to fulfill requests. Every time a user issues a request, an authentication procedure is performed to verify the user's identity against the metadata in their account.

\paragraph{User Registration.} When a new account is created, the user interacts with the Orchestrator layer to register the credentials they will use when interacting with the 0xPass platform, along with metadata on existing user accounts to be integrated into the universal account. User credentials may be obtained by linking to a third party single sign-on service already utilized by the user, or new credentials can be registered at the time of registration. In order to ensure strong security guarantees during later authentication phases, these credentials should include multiple authentication factors. Most importantly, the user will execute a protocol with the Orchestrator layer to generate a signing key to be stored by the user, while the corresponding verification key will be stored as account metadata and used to jointly authorize transactions later on. In summer, the user registration phase will create the following account information:

\begin{enumerate}
    \item Account metadata, including existing accounts information to be merged into the universal account.

    \item User credentials for authentication to the Orchestrator layer, either linked to a third party single sign-on service or newly registered by the user.

    \item User signing key for transaction authorization, to be stored locally by the user.
\end{enumerate}

\paragraph{Transaction Profiles.} Additionally, an account can be associated to a transaction profile determining which transactions can be authorized by the user through different types of authentication credentials and procedures. For example, This profile can impose transaction amount limits for individual transactions, for a given platform or for a given period of time. Alternatively, the profile can establish certain transaction patterns that are automatically allowed or denied. Each profile can be potentially associated to a given method to override the preferences in the profile and perform arbitrary transactions.

\paragraph{KYC and Identity Providers.} In the account registration procedure, the Orchestrator can optionally verify the user's identity via an existing Identity Provider. This process makes it possible to integrate Know Your Client (KYC) checks to the user's universal account and later connect this verified identity to individual accounts managed through the universal account. Support for different Identity Providers implementing multiple KYC processes can be added via new modules added to the Orchestrator layer.

\paragraph{Account Data Storage and Backup.} Once an account is created, all the metadata must be stored by the Orchestrator layer for authenticating users, solving requests and authorizing transactions. In order to provide fault tolerance and seamless recovery in case of outages, the Orchestrator layer can use modules executed by different sub-networks to store and/or backup such account data in multiple locations/platforms. In particular, these modules can be used to integrate existing Data Availability layers into the 0xPass platform for this purpose.

\subsection{User Authentication and Transaction Authorization Procedure.} In the authentication procedure, the user first proves their identity to the Orchestrator layer using their authentication credentials. Next, in order to issue a request, the user further signs their request along with a nonce provided by the Orchestrator layer using the signing key corresponding to the verification key registered as part of their account. The Orchestrator layer relays the signed request to the Solver layer along with an attestation that the user has also successfully authenticated themselves using their previously registered credentials. The Solver layer (and later the Transaction layer) only answers to requests that have been signed by the user in this manner and authorized by the Orchestrator layer. This procedure realizes an Identification Protocol secure against reset attacks~\cite{EC:BFGM01} that proves that the user has issued the transaction using the signing key corresponding to the verification key stored as part of the account. Combined with the multi-factor authentication credentials, this identification procedure ensures that the user is recognized by the 0xPass platform and has actively requested a certain transaction (via the signing procedure). In summary, user authentication and transaction authorization proceed as follows:

\begin{enumerate}
    \item The user authenticates their identity towards the Orchestrator layer via multi-factor credentials either registered directly at the Orchestrator layer or provided by a third party single sign-on solution.

    \item Upon successful authentication, the Orchestrator layer samples a random nonce and send it to the user's device.

    \item The user-side software signs the requested operation along with the nonce and returns the signed request to the Orchestrator layer.

    \item If the user's signature on the request (together with the nonce) is valid, the Orchestrator layer adds its own signature to the request and forwards it to the Solver layer to determine the sequence of transactions.

    \item Upon receiving the sequence of transactions from the Solver layer, the Orchestrator layer forwards these transactions for authorization by the user, who now authorizes the entire sequence of transactions by signing along with the same nonce previously issued by the Orchestrator for this session.

    \item Upon receiving the authorized sequence of transactions concatenated with the nonce, if the signature is valid, the Orchestrator adds its own signature and forwards the sequence of transactions along with the user's signature to the Transaction layer. 

    \item The Transaction layer verifies the user's and the Orchestrator's signatures on the sequence of transactions and, if the signatures are valid, generates signature on each transaction using the appropriate key for each platform where the transactions are to be executed. After this, the Transaction layer stores the nonce associated to this sequence of authorized transactions and no longer accepts requests with the same nonce.

\end{enumerate}

\paragraph{Storing User's Universal Account Signing Keys.} Initially, it is expected that a user's universal account signing key will be stored in a device controlled by the user. As the decentralization plan advances and more features are added to the 0xPass network, it will be possible to take advantage of the network itself for storing and protecting these keys. In the Second Phase of decentralization, a Key Management sub-layer will be added to the Transaction layer, allowing for securely storing signing keys as secret shares independently from the nodes executing different signing services in the Transaction layer. This Key Management sub-layer can also be used to securely store a user's universal account signing key in such a way that it can be requested by the user's device when needed. Besides providing key storage, this approach can also be used to backup the key by storing it in different sub-networks running the Key Management sub-layer, which improves fault tolerance and availability in case individual sub-networks are inaccessible. Throughout the development of the 0xPass network, users will be given the choice of storing their universal account signing keys locally or relying on the Key Management sub-layer.

\paragraph{Protecting User's Universal Account Signing Keys with Enclaves.} In order to reduce exposure of the user's signing key, it is possible to use secure enclaves for storing user secret keys to be used in the Identification Protocol. For platforms with powerful enough enclaves, it is possible to integrate that with authentication from the Orchestration layer, only running the identification protocol (\textit{i.e.} signing request/transaction sequences concatenated with the nonce) when a user is already authenticated by the Orchestration layer (\textit{e.g.} via checking single sign-on credentials).

\paragraph{Delegating Transaction Authorization.} In order to ensure that all transactions ever executed by the 0xPass platform are indeed authorized by the user, the authentication and authorization protocol above requires a lot of interaction with the users. This interaction can be eliminated by a scheme to delegate the right to authorize the specific sequence of transactions determined by the Solver layer to the Orchestrator layer or the Solver layer itself. This delegation allows for the Orchestrator layer or the Solver layer to directly authorize the sequence of transactions returned by the Solver layer ion behalf of the user and return to the Transaction layer for execution, while maintaining a high degree of user control over what transactions exactly can be authorized. In order to do so, the delegated transaction authorization solution uses Proxy Signatures~\cite{JC:BolPalWar12}, which allow for a signer to delegate the power to sign only certain messages to a third party (in this case the Solver or Orchestrator layers). The authentication and authorization protocol with delegation works as follows:

\begin{enumerate}
    \item The user authenticates their identity towards the Orchestrator layer via multi-factor credentials either registered directly at the Orchestrator layer or provided by a third party single sign-on solution.

    \item Upon successful authentication, the Orchestrator layer samples a random nonce and send it to the user's device.

    \item The user-side software signs the requested operation along with the nonce and returns the signed request to the Orchestrator layer. Moreover, the user-side software creates and sends to the Orchestrator layer a proxy signature certificate allowing the Solver layer to sign a sequence of transactions concatenated with the nonce provided by the Orchestrator layer such that the first transaction transfers the amount desired by the user from the source account specified by the user and that the last transaction transfers the amount desired by the user (within a given range to account for automatic exchanges and fluctuating rates) to the account specified by the user.

    \item If the user's signature and proxy signature certificate are valid, the Orchestrator layer adds its own signature to the request and certificate, creates a proxy signature certificate for the Solver layer to sign the same class of transactions on its behalf and forwards all to the Solver layer.

    \item Upon receiving a valid authorized request, the Solver layer uses the proxy signature certificates to create signatures on behalf of the user and the Orchestrator layer that authorize the sequence of transactions it has determined to be necessary to fulfill the request concatenated with the Orchestrator layer nonce, sending this authorized sequence of transactions to the Transaction layer. 

    \item The Transaction layer verifies the user's and the Orchestrator's signatures on the sequence of transactions and, if the signatures are valid, generates signature on each transaction using the appropriate key for each platform where the transactions are to be executed. After this, the Transaction layer stores the nonce associated to this sequence of authorized transactions and no longer accepts requests with the same nonce.

\end{enumerate}

\subsection{Account Recovery}

While public account metadata can be stored in a replicated manner over multiple modules executed over multiple sub-networks, private user credentials and the singing key must be closely guarded by the user. This requirement introduces risks of a user losing this crucial authentication information and being locked out of the system. If a single sign-on solution is used, the platform hosting this solution already implements account recovery procedures, mitigating the effect of credential loss. However, the signing key is a particularly vulnerable component of this system, since the user must store it locally in their device (or within a secure enclave in this device). We address these issues with procedures for securely backing up the signing key and for recovering a 0xPass universal account after irreversible credential loss.

\paragraph{Recovering the User's Signing Key.} One approach to backing up this signing key is simply storing secret shares of this key on different devices, including dedicated key management modules executed on specific sub-networks. This solution does not give access to the key to any individual device, but still requires a way for the user to prove their identity to these devices in order to recover the shares and reconstruct their key. The signing key can be more securely backed-up in one or more of the key management modules (executed on one of more sub-networks) using password-protected secret sharing~\cite{CCS:BJSL11,ACNS:JKKX17,JKKX16}. This procedure consists of splitting the secret key in shares that reveal no information unless a passkey is used when reconstructing the secret. In this setting, the password ensures that even if the devices reveal their shares to a user whose identity they cannot authenticate with high assurance, the shares can only be used to reconstruct the secret key by a user who possesses the right password. The password itself can be managed by current password management solutions already employed by the user, or stored in a physically secure location.

\paragraph{Recovering Account After Lost Credentials/Signing Key.} In case all credentials and/or the signing key are irreversibly lost, it is necessary to reset the user's authentication credentials and the signing key. This recovery procedure can be done automatically using long passphrases stored in a physically secure location or require careful off-chain ceremonies to verify a user's identity before allowing for a reset. These options are summarized bellow:

\begin{itemize}
     \item Establish off-chain account recovery mechanism and ceremonies for account recovery involving account recovery master keys kept in cold storage. This can be coupled with KYC process for off-chain user authentication if needed.

     \item Establish automatic account recovery using long passphrases that allow the user to reset their authentication credentials and signing key when presented to the Orchestrator layer. 
\end{itemize}

\section{Decentralization Phases}
At first, the 0xPass platform will operate as a centralized network fully controlled by the 0xPass organizations. This serves the purpose of bootstrapping the platform and stress testing the interoperability of each layer in the modular architecture. As the platform grows, it will undergo two decentralization phases, allowing third party organizations to run their own nodes in the 0xPass network as well as deploy their own implementations of new modules for the Solver Layer and for the Transaction Layer. At first, organizations wishing to join the platform or deploy new modules will need to be vetted in advance by the 0xPass organization, following a permissioned model. Later on, organizations will be able to have their nodes join the platform and deploy new modules in a permissionless way, staking different digital assets as collateral in the process. The decentralization phases are summarized as follows:

\begin{enumerate}
    \item \textbf{First Phase \-- Centralized Network Controlled by 0xPass.} In this phase, all layers are developed, deployed and controlled exclusively by 0xPass. Centralization will help deploy, test and quickly improve each module of all layers, which will pave the way for decentralizing the platform later. Although centralized, at no point any sensitive data (\textit{e.g.} signing keys) is entirely stored in any single node and users must actively authorize each transaction that is signed in a way that it is impossible for 0xPass to sign transactions without a users' consent.

    \item \textbf{Second Phase \-- Permissioned Decentralized Network.} In this phase, third party organizations will be able to join the 0xPass platform. The simplest way to join the platform will be running a node in the Solver and Transaction layers using 0xPass' software. Organizations that wish to deploy new features in the platform will also me able to implement new modules that will become available in the aforementioned layers. Regardless of how a third party organization joins the 0xPass platform, it will first be vetted by the 0xPass organization and enter contracts enforceable in a court of law.

    \item \textbf{Third Phase \-- Permissionless Decentralized Network.} In this phase, third party organizations will be able to deploy their own nodes executing existing software in the 0xPass platform or deploy new modules in a permissionless manner. The on-boarding of third party organizations will be automatically handled by the Orchestration layer, which will be augmented with a smart contract that will keep track of new third party organizations. In order to ensure user safety, third party organizations joining in this manner will be required to lock a certain amount of collateral funds, which will be used to reimburse users in case their nodes or software misbehaves or malfunctions.

\end{enumerate}

\section{First Phase \-- Centralized Network Controlled by 0xPass}

The main goal of the first phase is to establish a first version of all layers along with the basic infrastructure needed for interoperating across the first version Orchestrator, Solver and Transaction layers with a fixed set of modules. During this phase, all layers are fully controlled by 0xPass, who is responsible for deploying all modules and executing them on centralized sub-networks. This is necessary for ensuring a smooth bootstrapping of the system while allowing for easy deployment of new features and fast debugging of potential issues.

\paragraph{Basic Layered Architecture.} In this phase, the most basic version of the layered architecture is defined by means of an API for layer interoperation. This API defines the basic functions for layers to issue requests to each other and provide responses. At this point, the authorization of requests among layers and verification of authorization of requests and transactions by users is abstracted away, since the entire platform is controlled by 0xPass.

\paragraph{Orchestrator Layer.} The Orchestrator layer will implement a simple user registration and authentication scheme, with potential support for a handful of single sign-on platforms. At this stage, since all layers are executed by the 0xPass organization, user requests issued to the Orchestrator layer after successful user authentication are directly forwarded to the Solver and Transaction layers for execution without the need for direct authorization by the user. The Orchestrator layer directly authorizes requests received from the user and sequences of transactions determined by the Solver layer. Moreover, all account data is stored by the 0xPass organization.

\paragraph{Solver Layer.} In the First Phase, the Solver layer implements simple features for interoperability between a handful of cryptocurrency platforms. Support for more complex operations will be added as the 0xPass platform evolves.

\paragraph{Transaction Layer.} The first version of the Transaction layer supports threshold signing for the ECDSA signature scheme, with support to threshold Schnorr signing to be added as the system evolves.

\section{Second Phase \-- Permissioned Decentralized Network}
This phase will lay down the full modular layered architecture allowing for third party organizations to deploy modules and execute their own sub-networks. At this point, only third party organization vetted by the 0xPass organization according to a well-established process will be allowed to join the 0xPass network. It is expected that new modules implementing new features will be deployed by such third party organizations, which will be incentivized by means of a developer program. In particular, this phase will add a Key Management sub-layer to the Transaction layer where key shares can be backed-up in different sub-networks.

\paragraph{Permissioned Layered Architecture.}
In this phase, the layered architecture API will mature to allow for deploying third party modules and for executing modules on third party sub-networks. As part of this process, full support to the authentication and authorization flow (including delegation) will be added, which will allow sub-networks managed by different organizations and new modules to interoperate without jeopardizing security. Moreover, 0xPass will provide reference software for existing modules that can be executed over third party sub-networks for increased redundancy and network reliability.

\paragraph{Orchestrator Layer.} The Orchestration layer will be augmented to support the full authentication and authorization flow needed to issue requests and authorize transactions across all layers regardless of the organizations executing each module on different sub-networks. Support for more single sign-on services and KYC identity providers will be added in order to cater for a larger user base.

\paragraph{Solver Layer.} The Solver layer will be extended to handle more complex operations. Moreover, through the support for third party modules, it will be possible to add support for existing services that perform Solver layer duties. This integration will potentially be done by third party developers (after a vetting process).

\paragraph{Transaction Layer.} The Transaction layer will be extended to support new threshold signing schemes and new cryptocurrency platforms. Third party developers will be able to create new modules adding such features as needed (after a vetting process). Moreover, there will be the alternative to store signing key shares in a Key Management sub-layer executed by multiple sub-networks for increased fault tolerance.

\paragraph{Vetting Third Party Organizations.}
At this phase, 0xPass will define and streamline administrative and legal processes for vetting and on-boarding third party organizations. In order for an organization to deploy their own sub-network for executing existing modules in the 0xPass network, they must first demonstrate that they have the technical capacity to deploy and administer such services. Furthermore, due to the financially sensitive nature of this service, the organization must demonstrate that they have financial reserves to cover potential losses incurred by technical issues on their side. All of these financial guarantees and a technical Service Level Agreement (SLA) will be furthermore legally enforced by means of a contract to be entered by the 0xPass organization and any third party organizations who wish to deploy their own nodes and sub-networks at this phase.

\paragraph{Vetting Third Party Modules.} In order for a third party organization to deploy a new module, the software will first undergo an extensive audit  by reputable auditing organizations, followed for a period of testing on a testnet. This process aims at mitigating system-wide failures that could potentially be caused by a faulty module. 

\paragraph{Developer Program.} A pilot developers program will be started to incentivize third party developers and organizations to implement and deploy new modules and sub-networks. These program will give access to reasonable financial and technical resources to developers who wish to implement important new features in the 0xPass network by means of new modules.

\subsection{Sub-Networks and Key Management Sub-Layer.} 
As a step towards full decentralization, vetted third parties will be allowed to connect their own sub-networks of nodes to the 0xPass network in order to provide both new services via new modules and better fault tolerance when executing existing modules. These sub-networks will be integrated into the 0xPass network via the Authentication and Authorization mechanism, ensuring that all requests among sub-networks have been authorized by an authenticated user. 

\paragraph{Communication Across Sub-Networks.} In order to ensure that sub-networks running modules from different layers only execute user-approved requests, we will leverage the authentication and authorization mechanism described in Section~\ref{sec:auth}. This mechanism ensures that every request is authorized by a user whose identity has been successfully authenticated by the Orchestrator Layer. When receiving a request from any module in any layer, a module executed by a given sub-network first checks that this request has been authorized via the authentication and authorization protocol. If the request is properly authorized, the module answers the request. If not, the request is logged but denied. Besides verifying whether requests are authorized by authenticated users, all nodes in the 0xPass network communicate over secure channels established via the standard TLS 1.3 protocol, ensuring data privacy for both requests and the responses to requests. Moreover, this process guarantees that an audit trail is established for every request, allowing different layers and sub-networks to be audited later on (\textit{e.g.}, checking that a given request has been properly authorized by the relevant user).

\paragraph{Key Management Sub-Layer.} We will develop and deploy a pilot module for a new Key Management sub-layer that stores keys to be accessed by different modules in the Transaction layer. This module will be potentially also executed by sub-networks run by third party organizations on-boarded into the 0xPass network, which will improve the fault tolerance guarantees of the network as a whole. Following the authentication and authorization protocol from Section~\ref{sec:auth}, nodes in the Transaction layer can request signing key shares from nodes in the Key Management sub-layer when fulfilling a request by the Orchestrator or Solver layers. The Key Management sub-layer will verify that the request is authorized by an authenticated user and, if yes, transfer the signing key share assigned to each Transaction layer node directly to that node via a TLS 1.3 secure channel. Once the Key Management sub-layer is deployed, users will be able to store both their individual account signing keys and their universal account signing keys on multiple sub-networks executing Key Management modules. This will allow different Key Management sub-networks to act as backups for each other, as the Transaction layer will be able to obtain the necessary keys to sign a given transaction even if a given sub-network is momentarily inaccessible. Similarly, securely storing the universal account signing keys in multiple sub-networks will allow users to retrieve their key and authorize operations on the 0xPass network even if a given sub-network is temporarily inaccessible. 

\section{Third Phase \-- Permissionless Decentralized Network}
In this final decentralization phase, any third party organization can join the 0xPass network in an automated manner upon providing financial collateral to guarantee losses incurred by technical issues caused by this organization (\textit{i.e.} staking a certain amount of funds). The on-boarding, auditing and financial rewards/punishments of third party organizations will be managed by a smart contract connected to the Orchestrator layer. The ability to readily add new nodes or entire sub-networks and deploy new modules after an automated process will provide an easier route for organizations wishing the 0xPass platform. 

\paragraph{Permissionless Layered Architecture.}
In this phase, the layered architecture becomes open and permissionless, allowing any third party organization to deploy their own modules and run their own sub-network after an automated on-boarding process. The API will be extended to support a smart contract that will automate all the on-boarding a continuous auditing process for third party organizations. While third parties must communicate with this smart contract in order to join the network, this is transparent to the users and developers, whose experience remains exactly the same as in previous phases (\textit{i.e.} only having to interact with the Orchestrator layer via the same API).

\paragraph{Orchestrator Layer.} The Orchestrator layer is extended to interoperate with modules and sub-networks deployed by third party organizations that have been automatically on-boarded by means of a smart contract. Users and developers can choose whether they wish to use these third party modules and sub-networks to process requests, execute transactions and store signing key shares connected to their universal accounts.

\paragraph{Solver Layer.} The Solver layer operates as before but potentially gets new modules added by third party developers.

\paragraph{Transaction Layer.} The Transaction layer operates as before but potentially gets new modules added by third party developers. Moreover, the Transaction layer may use sub-networks from new third party organizations if the user's allow for their signing key shares to stored/processed in these sub-networks.

\paragraph{Smart Contract for Automatic On-Boarding and Auditing of Third Party Organizations.} In order to automatically on-board third party organizations without requiring any manual vetting or legally binding contracts, an Orchestration Smart Contract will be deployed in a suitable platform. The first role of this smart contract is locking collateral funds from each organization that joins the 0xPass network, so that this collateral can be used to automatically punish misbehaving organizations and reimburse users affected by malfeasance. After the on-boarding process, the smart contract monitors the execution of the modules hosted in sub-networks run by these third party organizations, also listening for complaints from other nodes and users. If a node or sub-network controlled by a given organization is proven to be misbehaving, the collateral funds are taken from this organization and distributed among honest ndoes and users who were affected by this misbehavior. This automated on-boarding, auditing and financial rewards/punishments can be implemented via the techniques in~\cite{FC:BauDavDow20,ACNS:BauDavFre21,FC:BCDF23}.

\paragraph{Key Management Sub-Layer with Rotating Committees.} As more nodes and sub-networks join the 0xPass network, the Key Management sub-layer can start using randomly selected subsets of parties (\textit{i.e.} random committees) to store signing key shares. These committees can be chosen publicly but they are particularly effective when chosen secretly, which protects the chosen nodes from DDoS and targeted attacks, since attackers do not exactly which nodes store each key shares. The YOSO model~\cite{C:GHKMNRY21} uses such secret rotating committees to achieve security against powerful adversaries while also improving scalability, since such committees can be significantly smaller than the total amount of nodes involved in the protocol execution. For the Key Management sub-layer we can use secret sharing and re-sharing protocols~\cite{AC:CDGK22} that allow anonymous random committees to transfer signing key shares among each other.

\bibliographystyle{plain}
\bibliography{additional,abbrev3,crypto}

\begin{thebibliography}{10}

\bibitem{CCS:BJSL11}
Ali Bagherzandi, Stanislaw Jarecki, Nitesh Saxena, and Yanbin Lu.
\newblock Password-protected secret sharing.
\newblock In Yan Chen, George Danezis, and Vitaly Shmatikov, editors, {\em ACM
  CCS 2011}, pages 433--444. {ACM} Press, October 2011.

\bibitem{FC:BCDF23}
Carsten Baum, James~Hsin{-}yu Chiang, Bernardo David, and Tore~Kasper
  Frederiksen.
\newblock Eagle: Efficient privacy preserving smart contracts.
\newblock In Foteini Baldimtsi and Christian Cachin, editors, {\em Financial
  Cryptography and Data Security - 27th International Conference, {FC} 2023,
  Bol, Bra{\v{c}}, Croatia, May 1-5, 2023, Revised Selected Papers, Part {I}},
  volume 13950 of {\em Lecture Notes in Computer Science}, pages 270--288.
  Springer, 2023.

\bibitem{FC:BauDavDow20}
Carsten Baum, Bernardo David, and Rafael Dowsley.
\newblock Insured {MPC}: Efficient secure computation with financial penalties.
\newblock In Joseph Bonneau and Nadia Heninger, editors, {\em FC 2020}, volume
  12059 of {\em {LNCS}}, pages 404--420. Springer, Heidelberg, February 2020.

\bibitem{ACNS:BauDavFre21}
Carsten Baum, Bernardo David, and Tore~Kasper Frederiksen.
\newblock {P2DEX}: Privacy-preserving decentralized cryptocurrency exchange.
\newblock In Kazue Sako and Nils~Ole Tippenhauer, editors, {\em ACNS 21,
  Part~I}, volume 12726 of {\em {LNCS}}, pages 163--194. Springer, Heidelberg,
  June 2021.

\bibitem{EC:BFGM01}
Mihir Bellare, Marc Fischlin, Shafi Goldwasser, and Silvio Micali.
\newblock Identification protocols secure against reset attacks.
\newblock In Birgit Pfitzmann, editor, {\em EUROCRYPT~2001}, volume 2045 of
  {\em {LNCS}}, pages 495--511. Springer, Heidelberg, May 2001.

\bibitem{JC:BolPalWar12}
Alexandra Boldyreva, Adriana Palacio, and Bogdan Warinschi.
\newblock Secure proxy signature schemes for delegation of signing rights.
\newblock {\em Journal of Cryptology}, 25(1):57--115, January 2012.

\bibitem{AC:CDGK22}
Ignacio Cascudo, Bernardo David, Lydia Garms, and Anders Konring.
\newblock {YOLO} {YOSO}: Fast and simple encryption and secret sharing in the
  {YOSO} model.
\newblock In Shweta Agrawal and Dongdai Lin, editors, {\em ASIACRYPT~2022,
  Part~I}, volume 13791 of {\em {LNCS}}, pages 651--680. Springer, Heidelberg,
  December 2022.

\bibitem{C:GHKMNRY21}
Craig Gentry, Shai Halevi, Hugo Krawczyk, Bernardo Magri, Jesper~Buus Nielsen,
  Tal Rabin, and Sophia Yakoubov.
\newblock {YOSO}: You only speak once - secure {MPC} with stateless ephemeral
  roles.
\newblock In Tal Malkin and Chris Peikert, editors, {\em CRYPTO~2021, Part~II},
  volume 12826 of {\em {LNCS}}, pages 64--93, Virtual Event, August 2021.
  Springer, Heidelberg.

\bibitem{JKKX16}
Stanislaw Jarecki, Aggelos Kiayias, Hugo Krawczyk, and Jiayu Xu.
\newblock Highly-efficient and composable password-protected secret sharing
  (or: How to protect your bitcoin wallet online).
\newblock In {\em {IEEE} European Symposium on Security and Privacy, EuroS{\&}P
  2016, Saarbr{\"{u}}cken, Germany, March 21-24, 2016}, pages 276--291. {IEEE},
  2016.

\bibitem{ACNS:JKKX17}
Stanislaw Jarecki, Aggelos Kiayias, Hugo Krawczyk, and Jiayu Xu.
\newblock {TOPPSS}: Cost-minimal password-protected secret sharing based on
  threshold {OPRF}.
\newblock In Dieter Gollmann, Atsuko Miyaji, and Hiroaki Kikuchi, editors, {\em
  ACNS 17}, volume 10355 of {\em {LNCS}}, pages 39--58. Springer, Heidelberg,
  July 2017.

\end{thebibliography}
\end{document}